**Quantum Key Distribution Without Authentication and Information Leakage**

Zixuan Hu[1] and Zhenyu Li[1,2]*

*[1]State Key Laboratory of Precision and Intelligent Chemistry, University of Science and Technology of China, Hefei 230026, China.*

*[2]Hefei National Laboratory, University of Science and Technology of China, Hefei 230088, China.*

**Email:* zyli@ustc.edu.cn

## ABSTRACT

Quantum key distribution (QKD) is the most extensively researched quantum cryptographic protocol, leveraging quantum phenomena to enable information-theoretically secure key establishment. Conventional QKD systems rely on authenticated classical post-processing channels to prevent impersonation attacks. However, QKD inherently lacks authentication capabilities, necessitating an external authentication mechanism. Furthermore, the classical post-processing steps introduce information leakage, exposing QKD to additional attack strategies and diminishing the final key rate. In this study, we introduce a novel QKD protocol that eliminates the need for separate authentication, prevents information leakage, and significantly improves the achievable key rate. By removing public classical channels entirely, our design attains perfect secrecy with the use of two reusable pre-shared keys.

## 1. Introduction

Quantum information science has made enormous progress over the last few decades [1-3]. One important branch of this field is quantum cryptography, which studies how quantum effects such as superposition, entanglement, and inherently probabilistic measurement can be utilized to design novel cryptographic models [4, 5]. The most prominent example among these is quantum key distribution (QKD) that provides information-theoretically secure key establishment and has been studied extensively with both rigorous theoretical models and state-of-the-art experimental implementations [6-11].

The conventional form of QKD (several forms have been shown to be equivalent) includes four classical post-processing steps: basis sifting (or simply “sifting”), parameter estimation (PE), error correction (EC), and privacy amplification (PA) [4, 5, 12]. All these steps are public and require entity authentication to ensure neither Alice nor Bob is impersonated by Eve (Alice is the sender, Bob the recipient, and Eve the adversary). Without authentication, Eve can launch man-in-the-middle (MITM) attacks by injecting or modifying classical messages during these steps and critically compromise the proved security of QKD [10, 13, 14]. Because QKD provides key establishment but no entity authentication, it relies on a separate authentication mechanism, which is cited as a core limitation and one of the main reasons for not recommending QKD in its current form for national security systems [15, 16]. In addition to its reliance on authentication, QKD also has information leakage in all classical post-processing steps that subjects it to additional classes of attacks (e.g. the delayed-choice attacks) and reduces its final key rate [4, 13, 17].

In this study, we propose a new QKD protocol that removes the need for a separate authentication mechanism, eliminates classical information leakage, and achieves a substantially increased key rate. We first present the new protocol that introduces two additional "protocol keys" as compared to conventional QKD. We then map the protocol to the mathematical model of the learning parity with noise (LPN) problem [18-20], and use this correspondence to prove information-theoretic security for the protocol, with the protocol keys being reusable. Next, we analyze the main advantages of the protocol over the conventional QKD model. Finally, tampering detection and error correction can be naturally and silently integrated into the new protocol for practical implementations.

## 2. Protocol Design

Consider the standard QKD setting where Alice and Bob share a quantum channel and a classical channel. Both channels are public that may be subject to Eve's tampering and examination. We assume that all three parties can perform arbitrary quantum operations and measurements on qubits. In conventional QKD [6, 7, 12], Bob has no intrinsic advantage over Eve with respect to the information carried by these channels: anything Bob can infer from the quantum and classical communications, Eve can infer as well. To securely distribute the key, Bob must therefore acquire an advantage via his interaction with Alice during the public classical post-processing steps. However, because Eve has the same capabilities as Bob, she can simply impersonate him in the classical communications, which is why an authentication mechanism is required. To remove the need for authentication, Bob must be able to perform some operations that Eve cannot emulate, without further assistance from Alice during the protocol. A natural way to achieve this is to let Alice and Bob pre-share additional secret keys. These secret keys would give Bob a private capability – shared only with Alice – that Eve does not possess, breaking the symmetry between Bob and Eve and allowing secure key establishment without authentication.

However, adding pre-shared secret keys to QKD introduces the burden of protecting them. Firstly, the pre-shared keys must be reusable, because otherwise the protocol will need to replenish them with the newly generated keys, effectively destroying the key rate. At the same time, the classical post-processing steps may leak information on the pre-shared keys, making them vulnerable when reused. Note that conventional QKD does not employ such keys and thus has no such a vulnerability, so a direct use of pre-shared keys will introduce new vulnerabilities to QKD that outweigh its potential benefits. Under these considerations, we need to introduce pre-shared keys in such a way that eliminates any information leakage that may compromise them, ensuring that these keys can be securely reused.

The first pre-shared key we have is a "basis key" that unifies the measurement bases for Alice and Bob – this gives Bob the advantage of always measuring in the same bases as Alice and therefore removes the sifting step that publicly compares the measurement bases. We acknowledge that such a basis key was already proposed in the Hwang-98 variation of QKD [21]. However, removing only the sifting step, Hwang-98 still has other public classical steps (PE, EC, and PA) that require authentication; in addition, the information leaked in these public steps can compromise the basis key when it is reused [22, 23]. For example, even without knowing the basis key, Eve can still guess the measurement bases and measure the qubits sent through the quantum channel: when Alice and Bob publicly compare the qubits' values in PE, Eve can verify if her guesses were correct and thereby infer partial knowledge on the basis key. So, to protect the basis key, PE and any other procedure that may reveal the qubits' values must be removed.

However, without PE, Alice and Bob cannot estimate how much Eve knows about the qubits' values, which makes PA (the process that converts the qubits' values into the final key that appears uniform to Eve despite her side information on the qubits' values) difficult because it relies on PE to determine how strongly the raw key must be compressed. If PA is not properly done, Eve's information on the qubits' values may translate into information on the final key, and QKD would lose its information-theoretic security. This interplay between the basis key, the qubits' values, the PE and PA processes, and the final key, motivates the following idea: if we make the correlation between the qubits' values and the final key a pre-shared secret between Alice and Bob, then Eve cannot relate her knowledge of the qubits' values to the final key – consequently PE and PA can be removed, the basis key can be safely reused, and the final key still has information-theoretic security. This pre-shared secret is precisely the 2nd pre-shared key we introduce in our new protocol: it is the crucial design element that allows us to eliminate the need for authentication and classical information leakage. Now the 1st pre-shared key is protected by the 2nd pre-shared key, we also need to prove the 2nd key's own security, which is done in Sec. 3 by mapping the protocol to the mathematical model of the LPN problem.

The details of our construction are described in Box 1. In the following discussion, the pre-shared keys used by our protocol itself are called the "protocol keys", while the key distributed by our protocol (to be used by other applications) is called the "application key". In Figures 1 and 2, we highlight the differences between conventional QKD and our new design.

**Box 1**

**Phase 0: pre-sharing of the protocol keys**

Step 0.1: Alice and Bob pre-share the 1st protocol key which is an $n$-bit string $\mathbf{h} = (h_1, \ldots, h_n)$.

Step 0.2: Alice and Bob pre-share the 2nd protocol key which is a $k \times n$ ($k < n$) matrix $\mathbf{F}$ with distinct rows $\mathbf{f}_1, \ldots, \mathbf{f}_k \in F_2^n$. The Hamming weights of all $\mathbf{f}_i$ and their nonzero linear combinations are not smaller than a security parameter $d$.

**Phase 1: generation and distribution of the application key**

Step 1.1: Alice creates $n$ EPR pairs of the form $\frac{1}{\sqrt{2}}(|00\rangle + |11\rangle)$.

Step 1.2: For each EPR pair, Alice picks one qubit $q_{Ai}$ to keep and the other one $q_{Bi}$ to be sent to Bob later.

Step 1.3: For each $q_{Bi}$ from $q_{B1}$ to $q_{Bn}$, Alice reads $h_i$, the $i$th bit of $\mathbf{h}$: if $h_i = 0$ she does nothing; if $h_i = 1$ she applies Hadamard to $q_{Bi}$.

Step 1.4: Alice sends each $q_{Bi}$ to Bob, in the order indexed by $i$.

Step 1.5: For each received $q_{Bi}$, Bob reads $h_i$: if $h_i = 0$ he does nothing; if $h_i = 1$ he applies Hadamard to $q_{Bi}$.

Step 1.6a: Alice measures all kept qubits in the current basis, obtaining $\mathbf{q}_A = (q_{A1}, \ldots, q_{An})^T$.

Step 1.6b: Bob measures all received qubits in the current basis, obtaining $\mathbf{q}_B = (q_{B1}, \ldots, q_{Bn})^T$. (In practice error correction can be added here)

Step 1.7a: Alice locally applies $\mathbf{F}$ to $\mathbf{q}_A$, generating a $k$-bit string $\mathbf{k}_A = \mathbf{F}\mathbf{q}_A$.

Step 1.7b: Bob locally applies $\mathbf{F}$ to $\mathbf{q}_B$, generating a $k$-bit string $\mathbf{k}_B = \mathbf{F}\mathbf{q}_B$.

Step 1.8: Ideally $\mathbf{k}_A = \mathbf{k}_B$, which then becomes the application key to be used in cryptographic applications.

For each Phase 0, Phase 1 can be repeated many rounds to generate many application keys with the same protocol keys reused. The number of repetitions allowed is discussed in Sec. 3.

Tampering detection happens naturally and silently in Step 1.8, when the application key is used: for details see Appendix C.

(Details of how to choose the parameters in the above protocol are given in Appendix A)

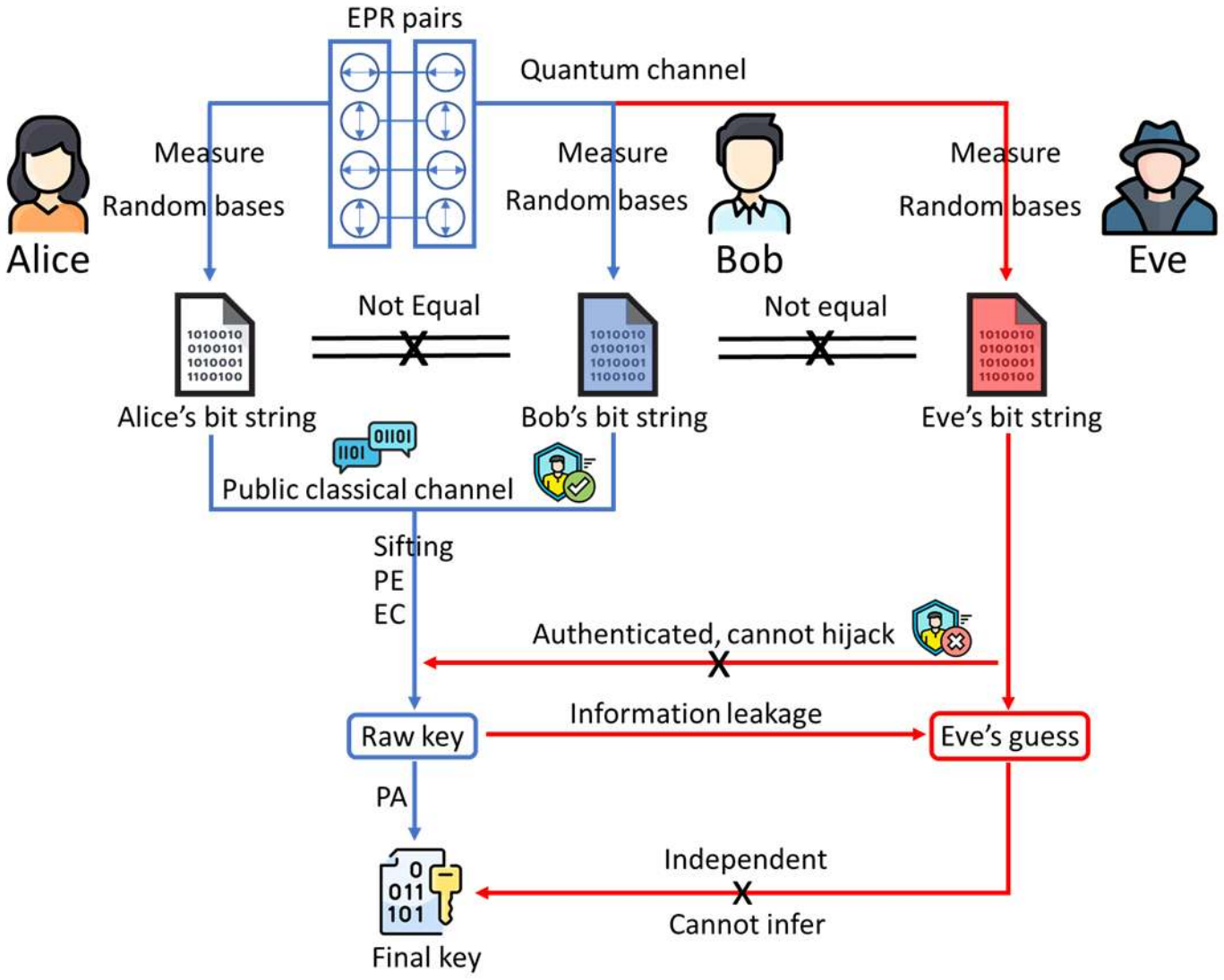

Figure 1. **Conventional QKD protocol.** Without any pre-shared key, Bob has the same capabilities as Eve when measuring the qubits and processing the resulting bit strings. Bob therefore relies on the public classical communications (sifting, parameter estimation (PE), and error correction (EC)) with Alice to post-process the measured bit strings into the final key. These communications require authentication to prevent Eve from impersonating Bob and running the post-processing herself with Alice. Information leakage during these communications also makes privacy amplification (PA) necessary, which substantially reduces the length of the final key.

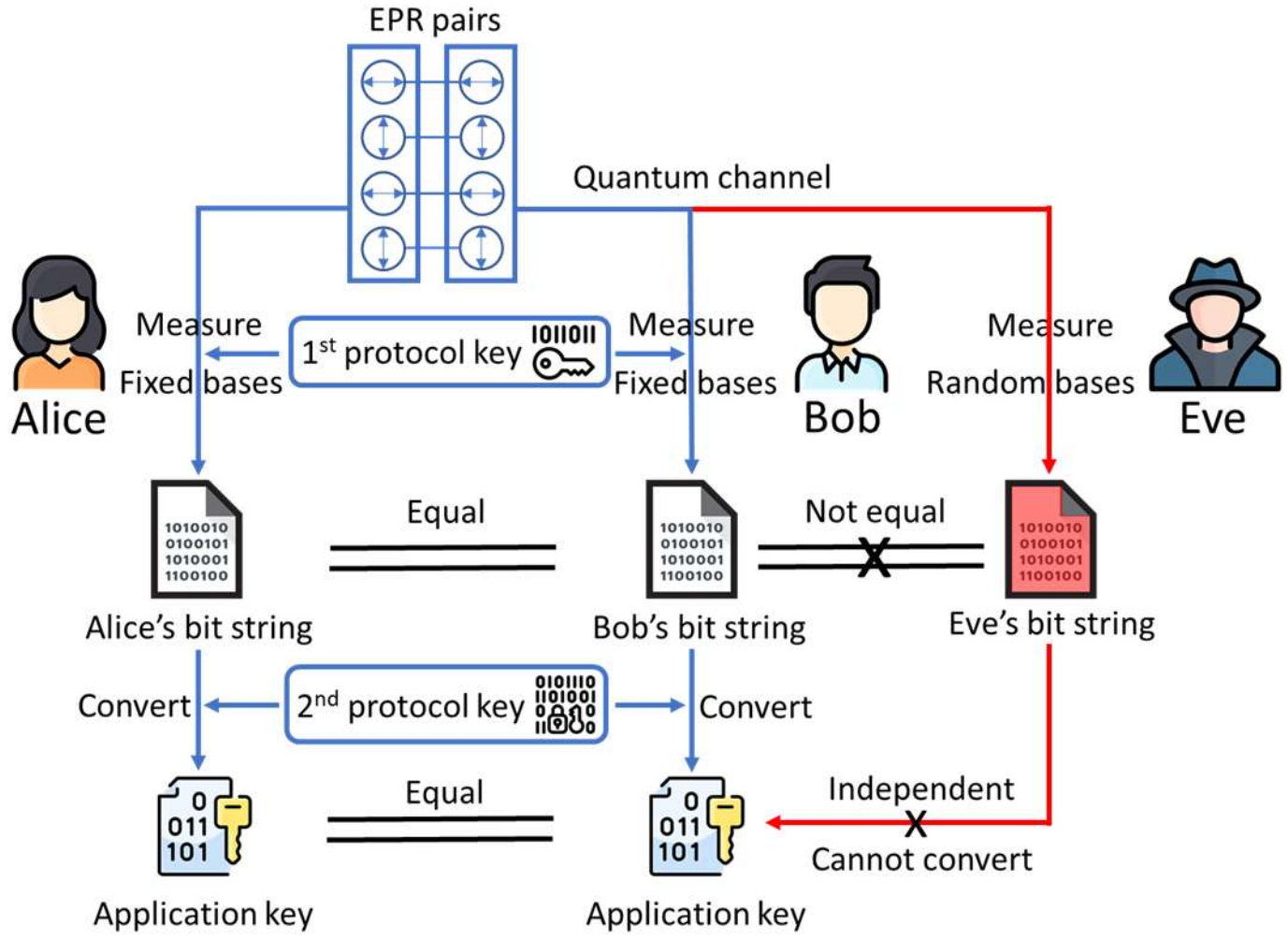

Figure 2. **New protocol design.** The 1st protocol key (pre-shared) unifies the measurement bases for Alice and Bob, so Bob already has an advantage over Eve when measuring the qubits. The 2nd protocol key (pre-shared) allows Alice and Bob to locally convert the measured bit strings into the application key: there is no public classical communications needed. Without the 2nd protocol key, Eve cannot relate her measured bit string to the application key by any means. Removing the public classical communications means the associated authentication and information leakage are also eliminated, greatly simplifying the protocol and increasing the final key length.

## 3. Mapping to the Learning Parity with Noise (LPN) Problem

To prove the security of our protocol in Box 1, we first show the addition of **F** with its associated procedures (Steps 1.6 through 1.8) allows us to map the new design to the mathematical model of the LPN problem [18-20].

**Definition of the LPN problem:**

Given a secret bit string $\mathbf{s}=(s_1,\ \ldots,\ s_n)\in F_2^n$ and noise $0<\tau<\frac{1}{2}$, suppose Eve observes samples of the pair $\left(\mathbf{a}^{(i)},\ b^{(i)}\right)=\left(\mathbf{a}^{(i)},\ \left\langle \mathbf{a}^{(i)},\mathbf{s}\right\rangle \oplus \varepsilon^{(i)}\right)$, where $i$ is the round number of the observations, $\left\langle \mathbf{a}^{(i)},\mathbf{s}\right\rangle=\sum_{j=1}^{n} \mathrm{a}_j^{(i)} s_j$, each $\mathbf{a}^{(i)}=\left(\mathrm{a}_1^{(i)},\ \ldots,\ \mathrm{a}_n^{(i)}\right)$ is uniformly distributed over $F_2^n$, each $\varepsilon^{(i)}$ is a Bernoulli process $\mathrm{Ber}(\tau)$. The goal of Eve is to find **s** given enough samples of $\left(\mathbf{a}^{(i)},\ b^{(i)}\right)$.

To map our design to the LPN problem, we note that:

$$\mathbf{k}_A=\mathbf{F}\mathbf{q}_A=\left(\left\langle \mathbf{f}_1,\mathbf{q}_A\right\rangle,\ \ldots,\ \left\langle \mathbf{f}_k,\mathbf{q}_A\right\rangle\right)^T \tag{1}$$

To know the application key, Eve needs to find both $\mathbf{q}_A$ and each $\mathbf{f}_j$. Without knowing the 1st protocol key **h**, Eve can only guess the measurement basis for the qubits intended for Bob, and thus she measures each $q_{iB}$ with an "error probability" $p$, resulting in her own $\mathbf{q}_E$ with the relation:

$$\mathbf{q}_E \oplus \mathbf{e}=\mathbf{q}_A \tag{2}$$

where $\oplus$ is bit-wise addition modulo 2, and each entry of $\mathbf{e}=(e_1,\ldots,e_n)^T$ is an independent Bernoulli process $\mathrm{Ber}(p)$. By Eqs. (1) and (2), the $j$-th entry of $\mathbf{k}_A$ is:

$$\mathbf{k}_{Aj}=\left\langle \mathbf{f}_j,\mathbf{q}_A\right\rangle=\left\langle \mathbf{f}_j,\mathbf{q}_E\oplus\mathbf{e}\right\rangle=\left\langle \mathbf{f}_j,\mathbf{q}_E\right\rangle\oplus\left\langle \mathbf{f}_j,\mathbf{e}\right\rangle \tag{3}$$

Then we establish the following identification:

$$\mathbf{q}_E=\mathbf{a}^{(i)},\quad \mathbf{k}_{Aj}=b^{(i)},\quad \mathbf{f}_j=\mathbf{s},\quad \left\langle \mathbf{f}_j,\mathbf{e}\right\rangle=\varepsilon^{(i)} \tag{4}$$

where the round number $i$ is omitted on $\mathbf{q}_E$, $\mathbf{k}_{Aj}$ and **e** for simplicity and consistency with Box 1, while it is understood that these variables are observed over multiple rounds. By Eq. (4) the new protocol is mapped to the LPN problem with a critical difference: LPN requires that Eve knows both $\mathbf{q}_E=\mathbf{a}^{(i)}$ and $\mathbf{k}_{Aj}=b^{(i)}$ to solve for $\mathbf{f}_j=\mathbf{s}$, while in Box 1 our protocol only reveals $\mathbf{q}_E$ to Eve but not $\mathbf{k}_{Aj}$: clearly whatever Eve does to obtain $\mathbf{q}_E=\mathbf{a}^{(i)}$ is entirely independent from the secret matrix **F** that Alice and Bob apply to their vectors, therefore our model is unsolvable for Eve and has stronger security than LPN.

Now if we weaken the protocol by allowing Eve to somehow apply known plaintext attacks [24] on some ciphertexts encrypted by $\mathbf{k}_{Aj}$ and obtain $\mathbf{k}_{Aj}$ for those rounds, then the problem of solving for $\mathbf{f}_j$ does reduce to LPN: in the following we show that even in this weakened scenario it still has almost perfect information-theoretic security in the sense that the information leakage per round is exponentially small in the security parameter *d*. The LPN problem is well known to be computationally hard if its noise term $\tau$ is bounded away from $\frac{1}{2}$, but becomes information-theoretically secure if $\tau$ approaches $\frac{1}{2}$ ([18], also see Appendix B for details). Now suppose the Hamming weight of one $\mathbf{f}_j$ is $w_j$, then the probability of $\langle \mathbf{f}_j, \mathbf{e} \rangle = \varepsilon^{(i)}$ being 1 gives the individual noise $\tau_j$ for $\mathbf{f}_j$:

$$\tau_j = \Pr\left(\langle \mathbf{f}_j, \mathbf{e} \rangle = 1\right) = \Pr\left(\sum_{i=1}^{n} f_{ji} e_i = 1\right) = \sum_{i\ odd} C\left(w_j, i\right) p^i \left(1-p\right)^{w_j - i} = \frac{1}{2}\left[1 - \left(1-2p\right)^{w_j}\right] \tag{5}$$

where *p* is the error probability of Eve measuring each entry of $\mathbf{q}_E$ different from $\mathbf{q}_A$, which is bounded away from both 0 and 1 due to ignorance of the 1st protocol key $\mathbf{h}$. Now define $\tau$ as our protocol's LPN-equivalent noise term, then because in Box 1 we have $w_j \geq d$ for all $\mathbf{f}_j$ and their nonzero linear combinations, Eve can never reduce $\tau$ below $\frac{1}{2}\left[1-\left(1-2p\right)^d\right]$. As $\tau$ approaches $\frac{1}{2}$ exponentially with *d*, our protocol has almost perfect information-theoretic security. With detailed derivation in Appendix B, the "success probability" $\gamma$ of Eve learning a single $\mathbf{f}_j$ correctly is related to the number *N* of the pairs of $\left(\mathbf{q}_E,\ \mathbf{k}_{Aj}\right) = \left(\mathbf{a}^{(i)},\ b^{(i)}\right)$ Eve needs to observe by the expression:

$$\gamma \leq N \cdot \left[1 - h_2\left(\tau\right)\right] \approx \frac{N}{2\ln 2}\left(1-2p\right)^{2w_j} \leq \frac{N}{2\ln 2}\left(1-2p\right)^{2d} \tag{6}$$

where $h_2\left(\tau\right) = -\tau \log \tau - \left(1-\tau\right)\log\left(1-\tau\right)$ is the binary entropy. Eq. (6) intuitively says Eve's success probability $\gamma$ is upper bounded by the product of the information leakage each round $\left[1-h_2\left(\tau\right)\right]$ and the total round number *N*: when $\left[1-h_2\left(\tau\right)\right] \to 0$ exponentially, *N* needs to be exponentially large, or else $\gamma$ would be exponentially small. To put Eq. (6) into concrete numbers, consider the intercept-resend attack where Eve measures each qubit in the Breidbart basis [4], by which she achieves the minimal error probability of $p = \frac{2-\sqrt{2}}{4}$: substituting this *p* into Eq. (6) we get:

$$\gamma \leq \frac{N}{\ln 2} \cdot 2^{-d-1} \tag{7}$$

Now if $d = 30$ and Eve is allowed to measure $\mathbf{q}_E$ and apply known plaintext attacks to get $\mathbf{k}_{Aj}$ for a million ( $N = 10^6$ ) repetitions of Phase 1 of Box 1, then she will guess a single $\mathbf{f}_j$ correctly with probability $\gamma \leq 0.07\%$. This extremely low probability will further decay exponentially with *d* increased; similarly, for any fixed success probability $\gamma$, *N* grows exponentially with the security parameter *d.*

**4. Security Analysis**

In the original and stronger protocol in Box 1, the only material available to Eve is the qubits $q_{iB}$ sent to Bob in Step 1.4. Each one of $q_{iB}$ is one half of a distinct EPR pair, so it is in the maximally mixed state, and the total density state of all $q_{iB}$ to Bob is:

$$\rho_B = \frac{1}{2^n} I \tag{8}$$

In Eq. (8) the identity matrix *I* is invariant upon arbitrary unitary transformation, so no matter what the 1st protocol key **h** is, $\rho_B$ remains the same. This means Eve can gain zero information on **h** by measuring $q_{iB}$ only, without accessing Alice's qubits $q_{iA}$. Note that this invariance of $\rho_B$ and zero information leakage hold even when the same **h** is reused arbitrarily many times. The 2nd protocol key **F** is clearly independent from the $q_{iB}$, so Eve gains zero information on **F** too, regardless of how many times **F** is reused.

For the weakened protocol described in Sec. 3, in addition to $\rho_B$, Eve also gains access to many rounds of $\mathbf{k}_{Aj} = \langle \mathbf{f}_j, \mathbf{q}_A \rangle$ through known plaintext attacks. However, without knowing $\mathbf{f}_j$, she cannot gain any information on $\mathbf{q}_A$ to compare with her measurement result $\mathbf{q}_E$, and would still have zero information on **h** regardless of how many times it is reused. On the other hand, as described in Sec. 3 and detailed in Appendix B, each round of $\mathbf{k}_{Aj}$ carries exponentially small amount of information ($\left[1 - h_2(\tau)\right]$) on the row $\mathbf{f}_j$ of **F**, when $\tau$ approaches $\frac{1}{2}$ exponentially: this is what we meant by having almost perfect information-theoretic secrecy. Note the total round number *N* in Eq. (6) sets an upper limit for how many times **F** can be reused: when this limit is approached, Alice and Bob can simply share a new **F** by a message protected by the application key – this resets the LPN problem and the limit of reuse. Assume that *k* bits of application key can be distributed per round, then about $k \cdot N$ bits can be distributed for close to *N* rounds. In the meanwhile we need $k \cdot n$ bits to encrypt the new $k \times n$ matrix **F**. For any fixed success probability $\gamma$, *N* grows exponentially with the security parameter *d*, so for even a moderately large *d*, $k \cdot N \gg k \cdot n$ and there are more than sufficient application key bits for updating **F**.

An important feature of our protocol is while **h** and **F** are reused, the application key $\mathbf{k}_A$ is randomly generated each round and can be used as the secret key for a one-time pad (OTP) encryption. Since the OTP encryption has perfect secrecy (zero information yielded to the adversary) [25], the protocol coupled with an OTP has perfect secrecy with the protocol keys **h** and **F** reusable, which violates Shannon's theorem that says perfect secrecy can only be achieved when the key is never reused [25]! Note that here the classical part of OTP still has the key $\mathbf{k}_A$ never reused, but the remarkable thing is the quantum part of the protocol can randomly generate a fresh $\mathbf{k}_A$ each round with the protocol keys reused while leaking no information – so Shannon's perfect-secrecy theorem is violated not in its original sense but in a new quantum sense [26]. Indeed, classically the key operates on the plaintext to produce the ciphertext, so these three are all correlated. Then to achieve perfect secrecy where the ciphertext yields no information on either the plaintext or the key, the correlation must be concealed by the randomness of the key. In our quantum protocol however, the "ciphertext" consists of the qubits $q_{iB}$ which are in the same maximally entangled state, regardless of **h** and **F**. Because these mechanisms are quantum only, when coupled with an OTP, our protocol demonstrates unique quantum advantage over classical cryptosystems by achieving a goal impossible in the classical regime.

## 5. Advantages over Conventional QKD

The first major advantage of the new design is it does not require authentication, which has already been analyzed in previous sections.

Secondly, the new protocol leaks zero information, while conventional QKD leaks considerable information in all classical post-processing steps: sifting tells Eve if she has measured each qubit in the correct basis; PE tells Eve the values of the compared qubits, whether her tampering has been detected, and if not detected which part of the raw key will be used later; in EC Eve can gain partial information on the raw key by reading the error correction syndrome; PA reveals how the raw key is converted into the final key. There are three main consequences of the information leakage:

1. Information leakage allows Eve to manipulate the information in her favor (e.g. biasing the sifting process or falsifying the PE data) and thus launch more damaging attacks in the absence of authentication. Without information leakage, our protocol does not have this increased vulnerability.

2. The leaked information can be exploited by Eve to launch the "delayed-choice" attacks on conventional QKD [4, 17]. In this attack strategy, Eve can entangle each traveling qubit $q_{iB}$ intended for Bob with her own probing qubits $q_{iE}$, and then forward $q_{iB}$ to Bob, while keeping $q_{iE}$ in the quantum memory. Eve then waits for Alice and Bob to announce their basis choices, and use this information to customize her strategy of measuring her quantum memory to achieve reduced qubit error rate (QBER) and increased probability of guessing $q_{iB}$'s values. Depending on how Eve measures her quantum memory, the attack can be individual, collective, or joint (generally coherent). Obviously, this attack strategy relies on the basis choice information leaked in the sifting step, which does not exist in our protocol. Without this information, Eve cannot learn any knowledge of Bob's measurements by measuring $q_{iE}$ entangled to $q_{iB}$: this is guaranteed by the quantum no-communication theorem [1]. Consequently, our protocol is not subject to the delayed-choice attacks.

3. To achieve information-theoretic security, information leakage must be treated by mechanisms that shorten the distributed key – this is discussed in more details for the next advantage where key rate is considered.

The third major advantage of the new protocol is having greatly increased key rate. In conventional QKD there are two factors that reduce the length of the distributed key: 1. In sifting, on average half of all qubits are discarded because Alice and Bob have not measured in the same basis. 2. Key length reduction due to information leakage: the qubits whose values are revealed in PE must be discarded; PA must reduce the key length for Eve's side information on the raw key before EC and the information leaked by the EC syndrome [4, 13, 27]. Considering these factors and the quantum leftover hash lemma, the asymptotic key rate is [12, 27]:

$$\lim_{n\to\infty} R \approx \frac{1}{2}\left[1-2h_2\left(Q\right)\right] \tag{9}$$

where $Q$ is the QBER estimated in PE. Note there are two "copies" of $h_2\left(Q\right)$ in the factor $\left[1-2h_2\left(Q\right)\right]$: one copy accounts for Eve's side information on $\mathbf{q}_A$ before EC, and the other accounts for the EC leakage [27]. For example if we set $Q=0.05$ in Eq. (9), then the asymptotic rate is $\lim_{n\to\infty} R \approx 0.214$. In comparison, the key rate of our protocol without EC approaches unity in the asymptotic limit (see Appendix A for details). If we add an additional EC mechanism, the asymptotic rate becomes:

$$\lim_{n\to\infty} R' \approx 1-h_2\left(Q\right) \tag{10}$$

Note that in our protocol, Eve's side information on $\mathbf{q}_A$ provides zero information on the final key $\mathbf{k}_A$, so there is only one copy of $h_2\left(Q\right)$. If we set $Q=0.05$ again in Eq. (10), then the asymptotic rate is $\lim_{n\to\infty} R \approx 0.714$. So our protocol's rate is significantly higher than that of conventional QKD. Setting the rate in Eq. (9) to zero and solving for $Q$ gives $Q_{\max} \approx 0.11$, which is the highest QBER that conventional QKD can tolerate [12, 27]. Now if we set the rate in Eq. (10) to zero and solving for $Q$, we get $Q_{\max} \approx 0.5$, which is the highest QBER that our new protocol can tolerate (0.5 is actually the information-theoretic limit). Consequently, our protocol can tolerate much higher QBER than conventional QKD.

The new protocol does not have the public classical steps in QKD, but can still have tampering detection, error estimation, and error correction for practical applications. For details please see Appendix C.

## 6. Conclusion

In this work, we introduced a new QKD protocol that integrates two additional protocol keys into the conventional QKD framework and maps naturally to the LPN problem. We then used the LPN framework to prove that our protocol has perfect information-theoretic security in its original form (with the protocol keys reusable), and almost perfect information-theoretic security in the weakened form allowing for KPA (with the protocol keys reusable up to a limit scaling exponentially with the security parameter $d$). Compared to conventional QKD, the new protocol relies on no external authentication mechanism, eliminates information leakage, and achieves a substantially higher key rate. In addition, our protocol can have natural and silent tampering detection, error estimation, and error correction processes for practical implementations.

**Acknowledgments**

This work is supported by the Innovation Program for Quantum Science and Technology (2021ZD0303306) and the National Natural Science Foundation of China (22393913).

**Conflicts of Interest**

The authors declare no conflicts of interest.

**Data Availability Statement**

All data are available in the main text or the Appendices (available after the References).

## Appendix A: Parameter Choice of the New Protocol in Box 1

The protocol in Box 1 has three main parameters: the block size *n*, the application key length *k*, and the security parameter *d*. In Step 0.2 of Box 1 we require the Hamming weights of all $\mathbf{f}_i$ and their nonzero linear combinations are not smaller than the security parameter *d*, so the linear space spanned by these $\mathbf{f}_i$ is called an $[n,k,d]$ code in standard coding theory [28], with **F** being its generator matrix. By results of coding theory, *n* must be much greater than *d* for such a code to exist [28]. In the meanwhile, for every *n* qubits sent to Bob, *k* bits of the application key $\mathbf{k}_A = \mathbf{k}_B$ are distributed, so the key rate is $R = \frac{k}{n}$ (without error correction), which we want as high as possible. Hence for the parameter choice we first fix the minimal *d* to meet our security requirement, then choose *n* as large as is practically feasible (so that a large *k* and $R = \frac{k}{n}$ remains achievable), and finally take *k* to be maximal subject to the existence of an $[n,k,d]$ code – existing codes can either be found on public databases or generated by known procedures [29]. For example, if we set $d \geq 30$ and $n = 256$, then the code we find is $[256,147,30]$ with $R = \frac{k}{n} \approx 0.574$; if $d \geq 30$ and $n = 512$, then the code we find is $[512,385,30]$ with $R = \frac{k}{n} \approx 0.752$ : a general rule is when *d* is fixed, *R* increases with *n* and approaches unity in the asymptotic limit ($n \to \infty$). Once *n*, *k*, and *d* are fixed, the *n*-bit string **h** can be randomly generated and pre-shared. Now for **F**, there are not many choices of $[n,k,d]$ codes, especially when we want to maximize *k*, so the code itself is not the source of entropy for **F** and can be publicly known. Instead, we note the rows $(\mathbf{f}_1,\ldots,\mathbf{f}_k)$ of **F** form an ordered basis of the *k*-D linear space of the $[n,k,d]$ code, so **F** belongs to the general linear group $GL(k,2)$ (the group of $k \times k$ invertible binary matrices) with the size $|GL(k,2)| = \prod_{i=1}^{k}\left(2^k - 2^{i-1}\right) = \Theta\left(2^{k^2}\right)$, therefore taking **F** to be a random member in $GL(k,2)$ provides about $k^2$ bits of entropy for this protocol key. In practice, the procedure of choosing **F** is the following: suppose we have already chosen an $[n,k,d]$ code, pick a basis of the code to form the rows of an initial generator matrix $\mathbf{F}_0$, generate a random invertible $k \times k$ binary matrix **M**, and then $\mathbf{F} = \mathbf{M}\mathbf{F}_0$ will serve as our pre-shared protocol key.

## Appendix B: Information-Theoretic Security of the Learning Parity with Noise (LPN) Problem and the New QKD Protocol

Consider the LPN problem defined in the main text Sec. 3, let $c^{(i)} = \left\langle \mathbf{a}^{(i)}, \mathbf{s} \right\rangle$, then $c^{(i)} = b^{(i)} \oplus \varepsilon^{(i)}$ , where $\varepsilon^{(i)}$ is a Bernoulli process Ber($\tau$) . Given the multiple-round knowledge of $\mathbf{A}_{1:N} = \left(\mathbf{a}^{(1)},...,\mathbf{a}^{(N)}\right)$ and $B_{1:i-1} = \left(b^{(1)},...,b^{(i-1)}\right)$, we can form the Markov chain $\mathbf{s} \to c^{(i)} \to b^{(i)}$, so by the data processing inequality we have:

$$I\left(\mathbf{s};b^{(i)} \mid \mathbf{A}_{1:N}, B_{1:i-1}\right) \le I\left(c^{(i)};b^{(i)} \mid \mathbf{A}_{1:N}, B_{1:i-1}\right) \qquad \text{B(1)}$$

where $I$ is the mutual information. By definition:

$$I\left(c^{(i)};b^{(i)} \mid \mathbf{A}_{1:N}, B_{1:i-1}\right) = H\left(c^{(i)} \mid \mathbf{A}_{1:N}, B_{1:i-1}\right) - H\left(c^{(i)} \mid b^{(i)}, \mathbf{A}_{1:N}, B_{1:i-1}\right) \le 1 - h_2\left(\tau\right) \qquad \text{B(2)}$$

where $h_2\left(\tau\right) = -\tau \log \tau - \left(1-\tau\right)\log\left(1-\tau\right)$ is the binary entropy and we have used the fact that $H\left(c^{(i)} \mid b^{(i)}, \mathbf{A}_{1:N}, B_{1:i-1}\right) = h_2\left(\tau\right)$ and $H\left(c^{(i)} \mid \mathbf{A}_{1:N}, B_{1:i-1}\right) \le 1$. Eq. B(2) intuitively means the information Eve learns about $c^{(i)}$ by observing $b^{(i)}$, given the knowledge of $\mathbf{A}_{1:N}$ and $B_{1:i-1}$, is upper bounded by $1 - h_2\left(\tau\right)$. Now by the chain rule of mutual information:

$$I\left(\mathbf{s};B_{1:N} \mid \mathbf{A}_{1:N}\right) = \sum_{i=1}^{N} I\left(\mathbf{s};b^{(i)} \mid \mathbf{A}_{1:N}, B_{1:i-1}\right) \le \sum_{i=1}^{N} I\left(c^{(i)};b^{(i)} \mid \mathbf{A}_{1:N}, B_{1:i-1}\right) \le N\left[1 - h_2\left(\tau\right)\right] \qquad \text{B(3)}$$

Also by the chain rule we have $I\left(\mathbf{s};B_{1:N}, \mathbf{A}_{1:N}\right) = I\left(\mathbf{s};\mathbf{A}_{1:N}\right) + I\left(\mathbf{s};B_{1:N} \mid \mathbf{A}_{1:N}\right)$. Because $\mathbf{s}$ is independent from all $\mathbf{a}^{(i)}$, $I\left(\mathbf{s};\mathbf{A}_{1:N}\right) = 0$, so we have:

$$I\left(\mathbf{s};B_{1:N}, \mathbf{A}_{1:N}\right) = I\left(\mathbf{s};B_{1:N} \mid \mathbf{A}_{1:N}\right) \le N\left[1 - h_2\left(\tau\right)\right] \qquad \text{B(4)}$$

Eq. B(4) intuitively means by observing the values of $b^{(1)}$ through $b^{(N)}$ and $\mathbf{a}^{(1)}$ through $\mathbf{a}^{(N)}$, Eve's information gain on $\mathbf{s}$ is upper bounded by the number $N$ of rounds multiplied by her information gain each round.

Next define $\gamma$ as the "success probability" that Eve guesses $\mathbf{s}$ correctly by observing $b^{(1)}$ through $b^{(N)}$ and $\mathbf{a}^{(1)}$ through $\mathbf{a}^{(N)}$, then by Fano's inequality we have:

$$I\left(\mathbf{s};B_{1:N}, \mathbf{A}_{1:N}\right) \ge \log\left(\left|S\right|\right) - h_2\left(\gamma\right) - \left(1-\gamma\right)\log\left(\left|S\right| - 1\right) \qquad \text{B(5)}$$

In Eq. B(5) $|S| \gg 1$ is the number of possible $\mathbf{s}$, so $\log\left(\left|S\right| - 1\right) \approx \log\left(\left|S\right|\right)$, and combining with Eq. B(4) we have:

$$\gamma \log\left(\left|S\right|\right) - h_2\left(\gamma\right) \le I\left(\mathbf{s};B_{1:N}, \mathbf{A}_{1:N}\right) \le N\left[1 - h_2\left(\tau\right)\right] \qquad \text{B(6)}$$

Now by the mapping of the new QKD protocol to LPN and Eq. (5) in the main text Sec. 3, $\tau = \frac{1}{2}\left[1 - \left(1-2p\right)^{w_j}\right]$, where $w_j$ is the Hamming weight of the $j$-th row of the protocol key $\mathbf{F}$, and $p$ is the error probability of Eve measuring each entry of $\mathbf{q}_E$ different from $\mathbf{q}_A$. Let $\delta = \frac{1}{2}(1-2p)^{w_j}$ and expanding $h_2(\tau)$ around $\frac{1}{2}$:

$$1-h_2(\tau) = 1-h_2\left(\frac{1}{2}-\delta\right) = \frac{2}{\ln 2}\delta^2 + O(\delta^4) \approx \frac{1}{2\ln 2}(1-2p)^{2w_j} \le \frac{1}{2\ln 2}(1-2p)^{2d} \qquad \text{B(7)}$$

where the last inequality holds because $\mathbf{F}$ is a generator matrix of an $[n,k,d]$ code such that $w_j \ge d$. By Eq. (4) in the main text Sec. 3 and Appendix A, $\mathbf{s}$ is a random vector in the $k$-dimensional $[n,k,d]$ space, so $|S| = 2^k$. Then by B(6) and B(7) we have:

$$\gamma k - h_2(\gamma) \le \frac{N}{2\ln 2}(1-2p)^{2d} \qquad \text{B(8)}$$

Note the binary entropy $h_2(\gamma)$ satisfies:

$$h_2(\gamma) = \frac{1}{\ln 2}\left[-\gamma\ln\gamma - (1-\gamma)\ln(1-\gamma)\right] \le \frac{1}{\ln 2}(-\gamma\ln\gamma + \gamma) = \gamma\log_2\frac{e}{\gamma} \qquad \text{B(9)}$$

For any $c > 1$, if $\gamma \ge ce2^{-k}$ then:

$$h_2(\gamma) \le \gamma\log_2\frac{e}{\gamma} \le \gamma(k - \log c) \qquad \text{B(10)}$$

Combining B(8) and B(10) we have:

$$\gamma k \le \frac{N}{2\ln 2}(1-2p)^{2d} + h_2(\gamma) \le \frac{N}{2\ln 2}(1-2p)^{2d} + \gamma(k-\log c) \qquad \text{B(11)}$$

Eq. B(11) can be simplified to:

$$\gamma \le \frac{N}{2\ln 2\log c}(1-2p)^{2d} \qquad \text{B(12)}$$

To reach Eq. B(12) we have assumed $\gamma \ge ce2^{-k}$, if the assumption fails then $\gamma < ce2^{-k}$ so if we let $c = 2$, we have:

$$\gamma \le \max\left\{\frac{N}{2\ln 2}(1-2p)^{2d},\ e2^{-k+1}\right\} \qquad \text{B(13)}$$

By our discussion on the parameter choice of the $[n,k,d]$ code in Appendix A, we pick the minimal $d$ to meet our security requirement, then choose $n$ as large as is practically feasible such that a large $k$ and $R = \frac{k}{n}$ remains achievable: we had $[256,147,30]$ and $[512,385,30]$ as examples. So within our parameter choice, $k \gg d$ and thus $\frac{N}{2\ln 2}(1-2p)^{2d}$ is much greater than $e2^{-k+1}$, so we have:

$$\gamma \le \frac{N}{2\ln 2}(1-2p)^{2d} \qquad \text{B(14)}$$

Consequently we have indeed proved Eq. (6) in the main text: which means the new QKD protocol has almost perfect information-theoretic security even when Eve is allowed to apply known-plaintext attacks to learn the application keys.

## Appendix C: Tampering Detection, Error Estimation and Error Correction for Practical Applications

Without knowing the protocol key $\mathbf{h}$, Eve's tampering with the qubits sent to Bob will alter their states and increase QBER in $\mathbf{q}_B$ for Bob. The overall QBER $Q$ can be defined as the error probability of Bob measuring each entry of $\mathbf{q}_B$ different from $\mathbf{q}_A$, then by replacing $p$ by $Q$ in Eq. (5) in the main text, Bob will have probability $\tau = \frac{1}{2}\left[1-(1-2Q)^{w_j}\right]$ for calculating any bit in the application key $\mathbf{k}_B$ not equal to $\mathbf{k}_A$. So any small QBER in $\mathbf{q}_B$ will be greatly amplified in $\mathbf{k}_B$: then tampering will be easily detectable when classical plaintexts encrypted by $\mathbf{k}_A$ become unintelligible to Bob. Once tampering is detected, Bob can send a pre-shared secret text (unknown to Eve, can be reused) encrypted by part of the application key shared in the current round to Alice, who decrypts it and estimate $Q$. Then Alice and Bob can perform error correction on $\mathbf{q}_A$ and $\mathbf{q}_B$ according to $Q$, with the syndrome also encrypted by part of last round's application key. A major advantage of our protocol over conventional QKD is the tampering detection, error estimation, and error correction processes all happen under the cover of regular classical communications encrypted by part of the application key shared in the current or last round, such that Eve does not know they have happened or whether she has been detected. Of course, by encrypting error estimation and correction we need to sacrifice part of the application key, and for the 1st round we need to pre-share additional initial keys in Phase 0 in Box 1, but this is worth it because we can have all the benefits of zero information leakage as discussed in the main text, at the cost of a small key rate reduction. Indeed, having perfect secrecy in our protocol, error estimation does not need to be run each round, and can be performed only after tampering is detected naturally and silently through regular communications encrypted by the application key. The practical version of our protocol including additional tampering detection, error estimation, and error correction processes is shown below:

**Phase 0: pre-sharing of keys**

Step 0.1: Alice and Bob pre-share the 1st protocol key which is an $n$-bit string $\mathbf{h} = (h_1, \ldots, h_n)$.

Step 0.2: Alice and Bob pre-share the 2nd protocol key which is a $k \times n$ ($k < n$) matrix $\mathbf{F}$ with rows $\mathbf{f}_1, \ldots, \mathbf{f}_k \in F_2^n$. The Hamming weights of all rows $\mathbf{f}_i$ and their nonzero linear combinations are not smaller than a security parameter $d$: i.e. $\mathbf{F}$ is a generator matrix of an $[n, k, d]$ code.

Step 0.3: Alice and Bob pre-share the error estimation secret text $\mathbf{t}=\left(t_1,...,t_l\right)$, and the initial error correction (EC) key $\mathbf{c}^{(0)}=\left(c_1^{(0)},...,c_m^{(0)}\right)$. The lengths $l \ll k$ and $m \ll k$ are picked to be the smallest numbers that can allow us to effectively perform error estimation and EC.

**Phase 1: generation and distribution of the application key**

Step 1.1: Alice creates $n$ EPR pairs of the form $\frac{1}{\sqrt{2}}\left(|00\rangle+|11\rangle\right)$.

Step 1.2: For each EPR pair, Alice picks one qubit $q_{Ai}^{(j)}$ to keep and the other one $q_{Bi}^{(j)}$ to be sent to Bob later: here $j$ is the round number ($j$ is omitted in the main text for simplicity).

Step 1.3: For each $q_{Bi}^{(j)}$ from $q_{B1}^{(j)}$ to $q_{Bn}^{(j)}$, Alice reads $h_i$, the $i^{\text{th}}$ bit of $\mathbf{h}$: if $h_i=0$ she does nothing; if $h_i=1$ she applies Hadamard to $q_{Bi}^{(j)}$.

Step 1.4: Alice sends each $q_{Bi}^{(j)}$ from $q_{B1}^{(j)}$ to $q_{Bn}^{(j)}$ to Bob, in the order indexed by $i$.

Step 1.5: For each received $q_{Bi}^{(j)}$, Bob reads $h_i$: if $h_i=0$ he does nothing; if $h_i=1$ he applies Hadamard to $q_{Bi}^{(j)}$.

Step 1.6a: Alice measures all kept qubits in the current basis, obtaining $\mathbf{q}_A^{(j)}=\left(q_{A1}^{(j)},...,q_{An}^{(j)}\right)^T$.

Step 1.6b: Bob measures all received qubits in the current basis, obtaining $\mathbf{q}_B^{(j)}=\left(q_{B1}^{(j)},...,q_{Bn}^{(j)}\right)^T$.

Step 1.6EC: Alice and Bob compute the EC syndromes $\mathbf{s}_A^{(j)}=\mathbf{H}\mathbf{q}_A^{(j)}$ and $\mathbf{s}_B^{(j)}=\mathbf{H}\mathbf{q}_B^{(j)}$ ($\mathbf{H}$ is the parity check matrix of an error-correcting code). Alice sends $\mathbf{s}_A^{(j)}\oplus\mathbf{c}^{(j-1)}$ to Bob: here $\oplus$ is bit-wise addition modulo 2; $\mathbf{c}^{(0)}$ is pre-shared, while all following $\mathbf{c}^{(j-1)}$ is a part taken from $\mathbf{k}_A^{(j-1)}$ of the previous round. Bob corrects $\mathbf{q}_B^{(j)}$. For QBER smaller than a pre-determined threshold, $\mathbf{s}_A^{(j)}$ is shorter than $\mathbf{c}^{(j-1)}$, in which case we only use part of $\mathbf{c}^{(j-1)}$ to encrypt $\mathbf{s}_A^{(j)}$.

Step 1.7a: Alice applies $\mathbf{F}$ to $\mathbf{q}_A^{(j)}$, generating a $k$-bit string $\mathbf{k}_A^{(j)}=\mathbf{F}\mathbf{q}_A^{(j)}$.

Step 1.7b: Bob applies $\mathbf{F}$ to the corrected $\mathbf{q}_B^{(j)}$, generating a $k$-bit string $\mathbf{k}_B^{(j)}=\mathbf{F}\mathbf{q}_B^{(j)}$.

Step 1.8: Alice sends $\mathbf{t}\oplus\left(k_{A1}^{(j)},...,k_{Al}^{(j)}\right)$ to Bob, who estimates the QBER $Q$. If $Q$ is close to zero, then reserve $\mathbf{c}^{(j)}=\left(k_{A(l+1)}^{(j)},...,k_{A(l+m)}^{(j)}\right)$ as the EC key for the next round, and use the rest of $\mathbf{k}_A^{(j)}$ as the application key. If $Q$ is nonzero but smaller than our threshold, repeat Phase 1 with longer EC syndromes. If $Q$ is larger than our threshold, abort the protocol and change channels.

For each Phase 0, Phase 1 can be repeated to generate many application keys with the same protocol keys reused. The number of repetitions allowed is discussed in the main text Sec. 3.